\documentclass[aip, jcp, twocolumn, floatfix, superscriptaddress, reprint, 10pt]{revtex4-1}  
\usepackage{graphicx}
\usepackage[usenames]{color}
\usepackage{microtype}
\usepackage{amsmath}
\usepackage{amssymb}
\usepackage{xspace}
\usepackage{footnote}
\usepackage{relsize}

\begin{document}

\title{Fast-Forward Langevin Dynamics with Momentum Flips
}

\author{Mahdi Hijazi}

\affiliation{Laboratory of Computational Science and Modeling, IMX, \'Ecole Polytechnique F\'ed\'erale de Lausanne, 1015 Lausanne, Switzerland}

\author{David M. Wilkins}

\email{david.wilkins@epfl.ch}

\affiliation{Laboratory of Computational Science and Modeling, IMX, \'Ecole Polytechnique F\'ed\'erale de Lausanne, 1015 Lausanne, Switzerland}

\author{Michele Ceriotti}

\affiliation{Laboratory of Computational Science and Modeling, IMX, \'Ecole Polytechnique F\'ed\'erale de Lausanne, 1015 Lausanne, Switzerland}

\begin{abstract}
Stochastic thermostats based on the Langevin equation, in which a system is coupled to an external heat bath, are popular methods for temperature control in molecular dynamics simulations due to their ergodicity and their ease of implementation.
Traditionally, these thermostats suffer from sluggish behaviour in the limit of high friction, unlike thermostats of the Nos{\'e}-Hoover family whose performance degrades more gently in the strong coupling regime. 
We propose a simple and easy-to-implement modification to the integration scheme of the Langevin algorithm that addresses the fundamental source of the overdamped behaviour of high-friction Langevin dynamics: if the action of the thermostat causes the momentum of a particle to change direction, it is flipped back. This fast-forward Langevin equation preserves the momentum distribution, and so guarantees the correct equilibrium sampling.  It mimics the quadratic behavior of Nos{\'e}-Hoover thermostats, and displays similarly good performance in the strong coupling limit. 
We test the efficiency of this scheme by applying it to a 1-dimensional harmonic oscillator, as well as to water and Lennard-Jones polymers. The sampling efficiency of the fast-forward Langevin equation thermostat, measured by the correlation time of relevant system variables, is at least as good as the traditional Langevin thermostat,
and in the overdamped regime the fast-forward thermostat performs much better, improving the efficiency by an order of magnitude at the highest frictions we considered. 
\end{abstract}

\maketitle

\section{Introduction}\label{sec:introduction}

The problem of thermostatting a molecular dynamics (MD) simulation, so that it samples a constant-temperature ensemble, is an important one in atomistic modelling: the solution of Hamilton's equations, the procedure at the heart of MD, produces trajectories sampling the microcanonical ensemble,\cite{AllenTildesley,FrenkelSmit} while in order to mimic the constant-temperature conditions that are most suitable to reproduce the usual experimental conditions, one would like to sample the canonical ensemble.
A thermostatting algorithm should have several desirable properties: it should be ergodic, sample the correct ensemble, and have a conserved quantity to check on the quality of the integration. Furthermore, an ideal thermostat would not affect the dynamics of the system too much, and would be straightforward to implement. A variety of thermostats have been proposed over the years, each of which fulfils these criteria to a varying degree.\cite{langevinSchneider,adelman1981,vangunsteren1981,Andersen,berendsen,nose,hoover,chains,VelocityRescaling,bussi_vresc2,Thermostats}

The earliest thermostatting methods were based on stochastic dynamics, mimicking the effect of random collisions with a heat bath. These included Langevin dynamics,\cite{langevinSchneider,adelman1981,vangunsteren1981} which controlled the temperature of a system by using a combination of random noise and frictional force, and the Andersen thermostat,\cite{Andersen} in which the momenta of particles are periodically re-drawn from a Maxwell-Boltzmann distribution.
Both of these methods are ergodic and sample the correct ensemble, but neither of these early algorithms had a conserved quantity, and both disturbed the system's dynamics to some degree.

A number of alternatives that do not depend on random numbers were also proposed. Among these was the weak-coupling algorithm of Berendsen,\cite{berendsen} which removed the random noise term from the Langevin equation to leave only a damping force: this thermostat is very straightforward to implement, but does not sample the correct ensemble.\cite{morishita2000}
Perhaps the most popular family of methods is based on the Nos{\'e}-Hoover (NH) thermostat,\cite{nose,hoover} in which a system is extended to include extra degrees of freedom, such that when the entire extended system is sampled in the microcanonical ensemble, canonical sampling of the original system is achieved. 
NH methods have a well-defined conserved quantity, but the original formulation suffered from nonergodic behaviour. This was solved by coupling a system to a chain of thermostats, the so-called Nos{\'e}-Hoover chain (NHC) method.\cite{chains} The resulting thermostat is robust and widely used, but suffers from the drawback that it is relatively difficult to implement.

Stochastic thermostats have recently seen a resurgence in popularity: in particular, a number of algorithms based on the Langevin equation have been introduced, including a stochastic velocity rescaling (SVR) thermostat\cite{VelocityRescaling,bussi_vresc2},
and a revisited Langevin dynamics thermostat.\cite{Langevin} 
Both of these methods include a well-defined conserved quantity, which allows the integration timestep to be controlled. These methods have formed the basis of a number of thermostats for path integral molecular dynamics (PIMD)\cite{PIMD} simulations, using either the white-noise or the generalized Langevin equation.\cite{LangevinColoredNoise,alacarte,ceriotti2010,ceriotti2011}
However, one key disadvantage remains in the Langevin thermostat compared to the NHC: in the limit of strong coupling to the system, the interaction of the Langevin thermostat with the system leads to frequent changes in the direction of particle momenta, hindering their exploration of phase space and greatly decreasing the efficiency of the thermostat. Nos{\'e}-Hoover methods do not suffer from this problem, and are applicable over a much wider range of coupling strengths.

In this paper, we introduce the fast-forward Langevin (FFL) thermostat, a modification to the standard Langevin dynamics algorithm whose goal is to counteract this inefficient sampling: the key idea of this method is that whenever the action of the thermostat changes the direction of the momentum, it is flipped back towards its initial direction.
This gives dynamics that are essentially those of a local version of the velocity-rescaling thermostat.\cite{VelocityRescaling,bussi_vresc2} As well as allowing the system to explore phase space more quickly at high frictions, the FFL thermostat should have no effect on the performance at low friction.

The remainder of this paper is set out as follows: in Sec.~\ref{sec:theory} we introduce fast-forward Langevin dynamics in one dimension and describe three possible generalizations to higher dimensions; in Sec.~\ref{sec:results} we apply the proposed algorithms to the simple harmonic oscillator, to liquid water and to a Lennard-Jones polymer; in Sec.~\ref{sec:discussion} we discuss our results; and finally, in Sec.~\ref{sec:conclusions} we draw our conclusions. 

\section{Theory}\label{sec:theory}

\subsection{Standard Langevin Dynamics}\label{sec:langevin}

The time-evolution of a single particle undergoing Langevin dynamics is described by the equations,
\begin{subequations}	 \label{eq:langevin_equation}
\begin{align}
\dot{\boldsymbol{q}} & = \boldsymbol{p} / m, \\
\dot{\boldsymbol{p}} & =-\gamma\boldsymbol{p}+\boldsymbol{\xi} + \boldsymbol{f},
\end{align}
\end{subequations}
where $\boldsymbol{q}$ is the particle's position, $\boldsymbol{p}$ its momentum, $m$ its mass, $\gamma$ the friction coefficient, $\boldsymbol{\xi}$ a random force sampled from a Gaussian distribution and $\boldsymbol{f}$ the force derived from an external potential.
The Cartesian components $\xi_{i}$ of the random force are uncorrelated in time and among themselves, i.e. $\left\langle \xi_{i}(t) \xi_{j}(t')\right\rangle = 2 m \gamma k_{\rm B} T \delta_{ij} \delta(t-t')$, where $k_{\rm B}$ is Boltzmann's constant and $T$ is the temperature.

A molecular dynamics integrator using the Langevin equation to sample the canonical ensemble was proposed by Bussi and Parrinello.\cite{Langevin} 
This integrator is based on the expression for time-propagation of the phase space probability density $P(\boldsymbol{p},\boldsymbol{q};t)$,
\begin{equation}\label{eq:liouville_evolution}
P(\boldsymbol{p},\boldsymbol{q};t + \Delta t) = e^{-\hat{L}\Delta t} P(\boldsymbol{p},\boldsymbol{q};t),
\end{equation}
where $\hat{L}$ is the Liouville operator,
\begin{equation}\label{eq:liouvillian}
\hat{L} = \boldsymbol{f}(\boldsymbol{q})\cdot \frac{\partial}{\partial \boldsymbol{p}} + \frac{\boldsymbol{p}}{m} \cdot\frac{\partial}{\partial \boldsymbol{q}} - \gamma \left( \frac{\partial}{\partial \boldsymbol{p}}\cdot\boldsymbol{p} + \frac{m}{\beta}\frac{\partial^{2}}{\partial p^{2}} \right),
\end{equation}
and $\beta = 1 / k_{\rm B} T$ the reciprocal temperature.

For a Liouvillian that can be written as the sum $\hat{L} = \sum_{j=1}^{n} \hat{L}_{j}$, the Trotter approximation allows us to write,\cite{trotter,tuckerman,sexton}
\begin{equation}
e^{-\hat{L} \Delta t} \simeq \prod_{j=1}^{n} e^{-\hat{L}_{j} \Delta t/2} \prod_{j=n}^{1} e^{-\hat{L}_{j} \Delta t/2},
\end{equation}
where $e^{-\hat{L}_{j} \Delta t/2}$ stands for propagation of a system under the Liouvillian $\hat{L}_{j}$ for a timestep $\Delta t/2$.
The Liouvillian of Eq.~\eqref{eq:liouvillian} can be written $\hat{L} = \hat{L}_{q} + \hat{L}_{p} + \hat{L}_{\gamma}$, where,
\begin{subequations}
\begin{equation}
\hat{L}_{q} = \frac{\boldsymbol{p}}{m} \cdot\frac{\partial}{\partial \boldsymbol{q}},
\end{equation}
updates particle positions,
\begin{equation}
\hat{L}_{p} = \boldsymbol{f}(\boldsymbol{q})\cdot \frac{\partial}{\partial \boldsymbol{p}},
\end{equation}
updates momenta under the action of the external force and,
\begin{equation}
\hat{L}_{\gamma} = - \gamma \left( \frac{\partial}{\partial \boldsymbol{p}}\cdot\boldsymbol{p} + \frac{m}{\beta}\frac{\partial^{2}}{\partial p^{2}} \right),
\end{equation}
\end{subequations}
updates momenta under the action of the Langevin forces.

In Ref.~\onlinecite{Langevin} the splitting chosen for the propagator was,
\begin{equation}\label{eq:propagator_trotter}
e^{-\Delta t \hat{L}} \simeq e^{-\Delta t \hat{L}_{\gamma}/2} e^{-\Delta t \hat{L}_{p}/2} e^{-\Delta t \hat{L}_{q}} e^{-\Delta t \hat{L}_{p}/2} e^{-\Delta t \hat{L}_{\gamma}/2},
\end{equation}
although other splittings of the propagator are possible, and have been shown to give more accurate results for large time steps.\cite{leimkuhler2013}
When $\gamma = 0$,
Eq.~\eqref{eq:propagator_trotter} has the same form as the familiar velocity Verlet algorithm for NVE molecular dynamics,\cite{CompSimLiquids} making it very straightforward to implement.
Eq.~\eqref{eq:propagator_trotter} is applied to an MD simulation by evolving the particle momenta $\boldsymbol{p}$ and positions $\boldsymbol{q}$ through a timestep $\Delta t$ according to the scheme,
\begin{subequations}\label{eq:langevin_integrator}
\begin{align}
\boldsymbol{p} & \leftarrow c_{1}\boldsymbol{p} + c_{2}\boldsymbol{R},\label{eq:thermostat_1} \\
\boldsymbol{p} & \leftarrow  \hphantom{c_{1}}\boldsymbol{p} + \boldsymbol{f} \Delta t/2,\label{eq:verlet_1} \\
\boldsymbol{q} & \leftarrow  \hphantom{c_{1}}\boldsymbol{q} + \frac{\boldsymbol{p}}{m} \Delta t,\label{eq:verlet_2} \\
\boldsymbol{p} & \leftarrow  \hphantom{c_{1}}\boldsymbol{p} + \boldsymbol{f} \Delta t/2,\label{eq:verlet_3} \\
\boldsymbol{p} & \leftarrow c_{1} \boldsymbol{p} + c_{2}\boldsymbol{R}',\label{eq:thermostat_2}
\end{align}
\end{subequations}
where $\boldsymbol{R}$ and $\boldsymbol{R}'$ are two independent Gaussian random numbers with zero mean and unit variance and,
\begin{subequations} \label{eq:Langevin_constants}
\begin{align}
c_1 & =e^{-\gamma (\Delta t/2)}, \label{eq:damping} \\
c_2 & =\left[(1-c_1^2)\frac{m}{\beta}\right]^{1/2}. \label{eq:random_noise}
\end{align}
\end{subequations}
The constant $c_{1}$ accounts for damping of momenta,
and $c_{2}$ for the action of the random noise.
If $\gamma = 0$ there is no frictional force, and $c_{1} = 1$ and $c_{2} = 0$, so that the thermostat steps (Eqs.~\eqref{eq:thermostat_1} and \eqref{eq:thermostat_2}) have no effect and the integration scheme becomes the traditional velocity Verlet algorithm.
The integration timestep $\Delta t$ can be controlled using the conserved quantity described in Ref.~\onlinecite{Langevin}, which corrects the total energy with a counter that keeps track of the energy exchanged with the external heat bath.

In order to optimize the sampling efficiency of the Langevin thermostat, the friction parameter $\gamma$ must be tuned: if the friction is too low then the thermostat will have too little an effect on the system, and if it is too high, it will lead to overdamping and sluggish exploration of configuration space.
This inefficiency at high frictions is a disadvantage that is not shared by the NH thermostat.\cite{ceriotti2010} Here we will show that the difference stems essentially from the fact that the nonlinear NH equations cannot change the sign of the momentum. One can achieve the same effect with a simple modification of Langevin dynamics: whenever the action of the thermostat changes the sign of the particle momentum, it should be flipped back to the original direction. We will show that doing so eliminates the effects of overdamping.

\subsection{Fast-Forward Langevin Dynamics}

In a 1-dimensional system, flipping the sign of the momentum can be achieved straightforwardly by replacing Eq.~\eqref{eq:thermostat_1} with the following two steps,
\begin{align}\label{eq:new_thermostat_1}
p & \leftarrow c_{1} p_{0} + c_{2} R, \nonumber \\
p & \leftarrow \text{sign}(p p_{0}) p, \tag{\ref{eq:thermostat_1}$^{\prime}$}
\end{align}
and making a similar replacement for Eq.~\eqref{eq:thermostat_2} (with $p_{0}$ storing the value of the momentum before the thermostat step in each case).
We call the resulting algorithm the fast-forward Langevin (FFL) thermostat.
The flipping step in Eq.~\eqref{eq:new_thermostat_1} leaves the magnitude of the momentum unchanged;
because the canonical probability distribution for the momentum depends only on its magnitude, FFL dynamics generate momenta drawn from the correct distribution for this ensemble, although it is not ergodic and relies on the action of the potential to change the sign of the momentum.
One could also regard this algorithm as arising from the solution of a stochastic differential equation for $p^2$, that is then implemented as a scaling of the value of $p$ -- a point of view that might be better suited to study it analytically, and to reveal the connections to other thermostatting schemes such as NH and stochastic velocity rescaling.  

In multiple dimensions, this algorithm is not as straightforward to implement as in 1D, because the sign of the momentum cannot be unambiguously defined. This means that there are several possibilities for the definition of the FFL thermostat in higher dimensions; we will consider three of these possibilities.

\begin{figure}[hbt!]
\includegraphics[scale=0.25]{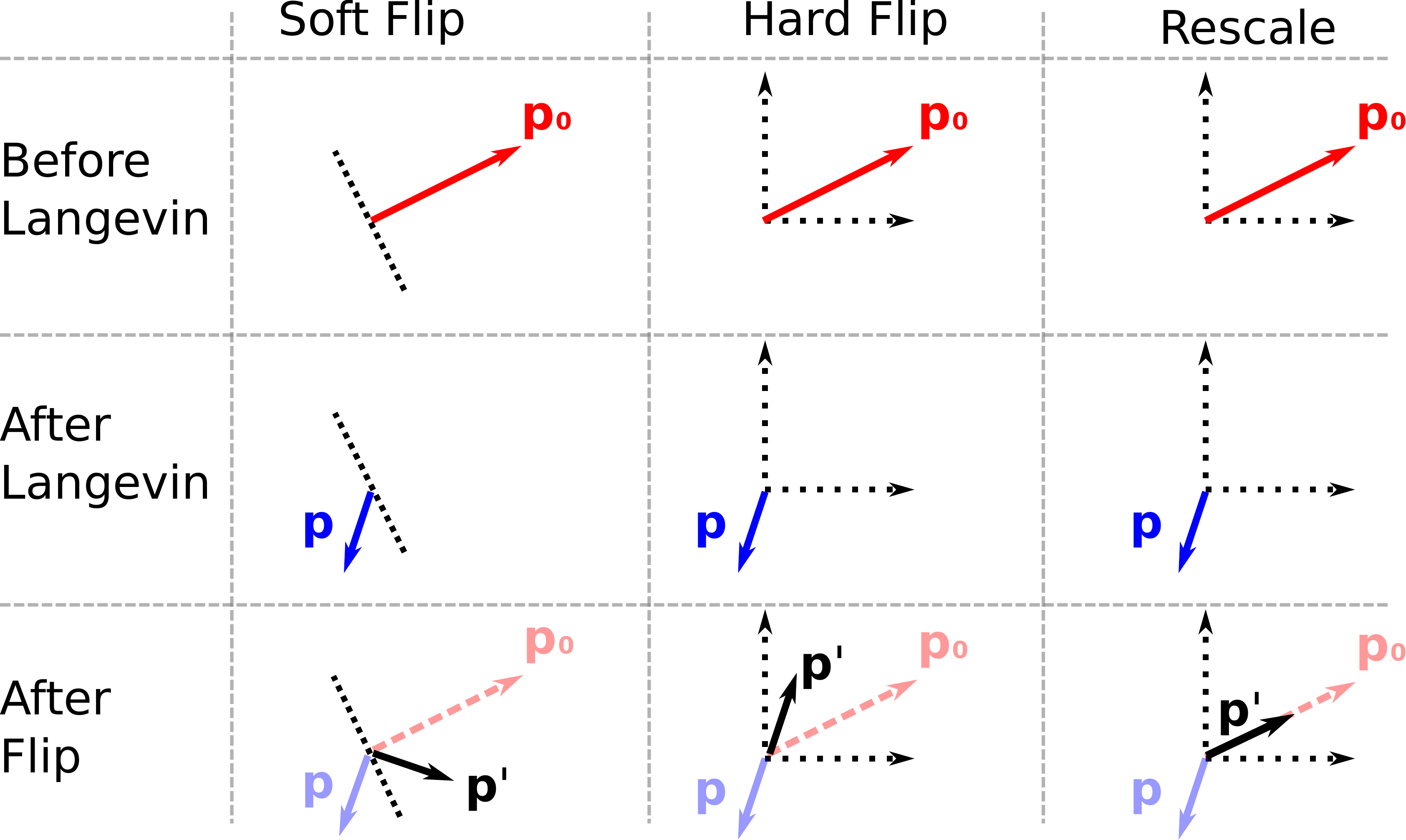}
\caption{\label{fig:algorithms} Illustration (in 2 dimensions) of the different flipping algorithms. $\boldsymbol{p}_{0}$ is the momentum before the thermostat step (Eq.~\eqref{eq:thermostat_1}), $\boldsymbol{p}$ is the momentum after this step, and $\boldsymbol{p}'$ is the momentum after the flip is performed. The soft flip reflects the momentum through the plane perpendicular to $\boldsymbol{p}_{0}$. The hard flip applies the 1-dimensional algorithm to each axis independently. The rescale flip preserves the initial direction of $\boldsymbol{p}_{0}$ but assigns it the magnitude of $\boldsymbol{p}$.}
 \end{figure}

In the first method, a particle's momentum changes sign whenever $\boldsymbol{p}_{0}\cdot\boldsymbol{p} < 0$: that is, the action of the thermostat moves the momentum into the opposite half-space to that in which it began. In this case, $\boldsymbol{p}$ is reflected through the plane perpendicular to $\boldsymbol{p}_{0}$. This algorithm, shown in the first column of Fig.~\ref{fig:algorithms}, is described as a ``soft'' flip. Eq.~\eqref{eq:thermostat_1} is replaced by,
\begin{align}
\boldsymbol{p} & \leftarrow c_1\boldsymbol{p}_{0}+c_2\boldsymbol{R}, \nonumber \\
\boldsymbol{p} & \leftarrow \boldsymbol{p} - 2 \frac{F\left[\boldsymbol{p}\cdot\boldsymbol{p}_{0}\right]} {\boldsymbol{p}_0\cdot\boldsymbol{p}_0}\boldsymbol{p}_0,
\tag{\ref{eq:thermostat_1}$^{\prime\prime}$}
\end{align}
where $F[x] = x$ if $x<0$, and $F[x]=0$ otherwise.

Rather than considering the momentum vector of a particle, the second method considers individual Cartesian components, and the 1D algorithm is applied in each dimension.
This ``hard'' flipping algorithm is shown in the second column of Fig.~\ref{fig:algorithms}. Eq.~\eqref{eq:thermostat_1} is replaced by Eq.~\eqref{eq:new_thermostat_1} in each Cartesian direction.

The hard flip algorithm leads to dynamics that depend on the choice of axis system. For isotropic systems this is not expected to lead to any problems, but in anisotropic systems it would be necessary to choose these axes appropriately, a problem that has also been observed for NHC thermostats.\cite{alacarte}
A more elegant, and rotationally invariant, algorithm that does not rely on the specific choice of axis system is
a ``rescale'' flip, shown in the third column of Fig.~\ref{fig:algorithms}: after the action of the thermostat, the direction of $\boldsymbol{p}$ is set to the direction of $\boldsymbol{p}_{0}$, while keeping the new magnitude. Eq.~\eqref{eq:thermostat_1} is thus replaced by,
\begin{equation}
\boldsymbol{p} \leftarrow \left| c_{1} \boldsymbol{p} + c_{2} \boldsymbol{R}\right| \frac{\boldsymbol{p}_{0}}{\left|\boldsymbol{p}_{0}\right|}. \tag{\ref{eq:thermostat_1}$^{\prime\prime\prime}$}
\end{equation}
This rescale flip, in which the thermostat evolves the magnitude of each particle's momentum without changing its direction, is
essentially equivalent to applying a SVR thermostat \cite{VelocityRescaling,bussi_vresc2} to each particle, crucially without applying a correction that allows particles to change sign.

\subsection{Thermostat Efficiency}

In order to compare the efficiency of two thermostats, we use the correlation times of physical observables.
The autocorrelation function of variable $A$ is,
\begin{equation}
C_{AA}(t) = \frac{\left\langle\delta A(t) \delta A(0)\right\rangle}{\left\langle (\delta A)^{2}\right\rangle},
\end{equation}
with $\delta A(t) = A(t) - \left\langle A \right\rangle$ is the instantaneous deviation of the variable from its average value.
For this correlation function we define a correlation time as,
\begin{equation}
\tau_{A} = \int_{0}^{\infty} C_{AA}(t)\,{\rm d} t.
\end{equation}
The smaller $\tau_{A}$ is, the faster the system is decorrelated and the more efficiently the variable is sampled by the thermostat.

We will express the strength of the thermostats in terms of their intrinsic relaxation times. For a Langevin thermostat, this is given by $\tau_0=1/\gamma$, whereas for a NHC thermostat we use the definition $\tau_0=\sqrt{Q\beta/4}$, where $Q$ is the thermostat mass.\cite{ceriotti2010}
 
\section{Results}\label{sec:results}

\subsection{Harmonic Oscillator}\label{sec:harmonic_oscillator}

\begin{figure}[h!]
\includegraphics[scale=0.5]{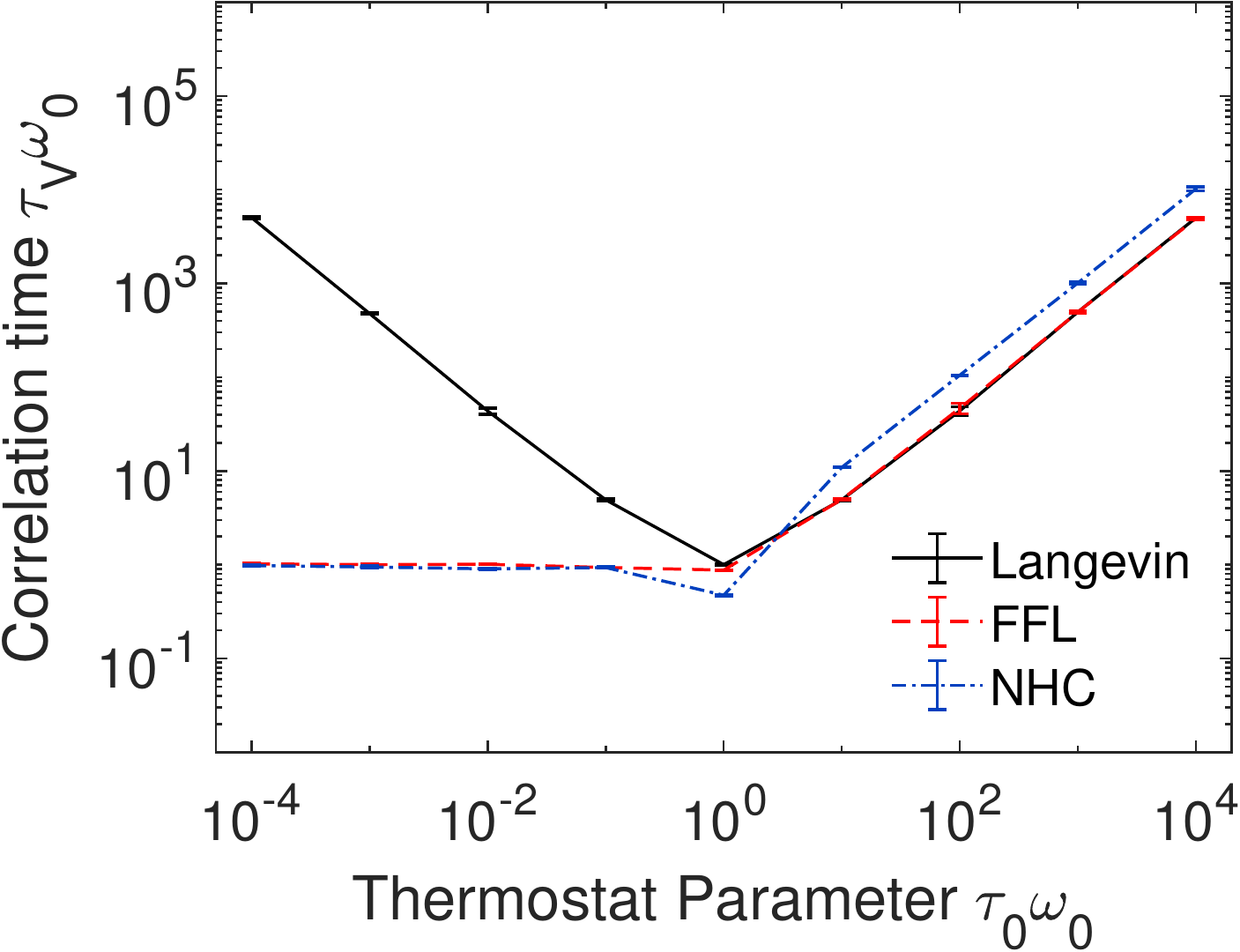}
\caption{\label{fig:Har} 
Potential energy autocorrelation time $\tau_{V}$ for a 1D harmonic potential with frequency $\omega_{0}$, for the standard Langevin thermostat, FFL thermostat and the NHC thermostat.
}
\end{figure}

The first system to which we applied the FFL thermostat was a 1D harmonic oscillator potential with frequency $\omega_0$ and potential energy,
\begin{equation}
V(q) = \frac{1}{2} m \omega_{0}^{2} q^{2}.
\end{equation}
In Fig.~\ref{fig:Har} we show the potential energy correlation time $\tau_{V}$ for the Langevin and FFL thermostats at different values of the characteristic time $\tau_{0} = 1/\gamma$, and for the NHC thermostat with $\tau_{0} = \sqrt{Q\beta/4}$, where $Q$ is the thermostat mass.\cite{chains,ceriotti2010} In each case, the target temperature was $T = \hbar \omega_{0} / k_{\rm B}$.
The correlation time for the harmonic oscillator coupled to a Langevin thermostat obeys the relation,\cite{zwanzig,gardiner}
\begin{equation}
\tau_{V} \omega_{0} = \frac{1}{2} \left( \tau_{0}\omega_{0} + \frac{1}{\tau_{0}\omega_{0}} \right),
\end{equation}
having a minimum when $\tau_{0}\omega_{0} = 1$ (so that $\gamma = \omega_{0}$).

On the other hand, for the NHC thermostat $\tau_{V}$ increases with $\tau_{0}$ when $\tau_{0} > 1/\omega_{0}$, but below this point the correlation time is essentially constant, meaning that this thermostat can be used efficiently over a much wider range of friction constants.
Comparing both of these thermostats to the FFL thermostat, we note that for low frictions it gives the same correlation time as the standard Langevin equation, when the action of the thermostat hardly ever changes the sign of the momentum and the fast-forward step has no effect. On the other hand, for high frictions the correlation time remains roughly equal to its optimum value: the flipping of the momenta clearly gives a great improvement in the exploration of phase space, and removes a major disadvantage of the Langevin thermostat compared to the NHC at high friction.
It should be stressed that in order to obtain meaningful results in the limit of very small $\tau_0$, one has to use a time step $\Delta t$ that is smaller than $\tau_0$.
If instead $\Delta t \gg \tau_{0}$, then Eqs.~\eqref{eq:Langevin_constants} show that $c_{1}$ and $c_{2}$ will saturate at their high-friction values and the effective coupling strength saturates at the value corresponding to $\Delta t$.
When simulating a single oscillator, there is no reason to use a time step which is much smaller than $1/\omega_0$. Results in the strongly overdamped regime $\tau_0\omega_0 \ll 1$ are however very relevant in real systems, which contain multiple vibrational modes spanning several orders of magnitude. In such a case, the time step is determined by the fastest modes in the system, while overdamping would also affect -- and be most detrimental to -- slower vibrations.

\begin{figure}[h!]
\includegraphics[scale=0.5]{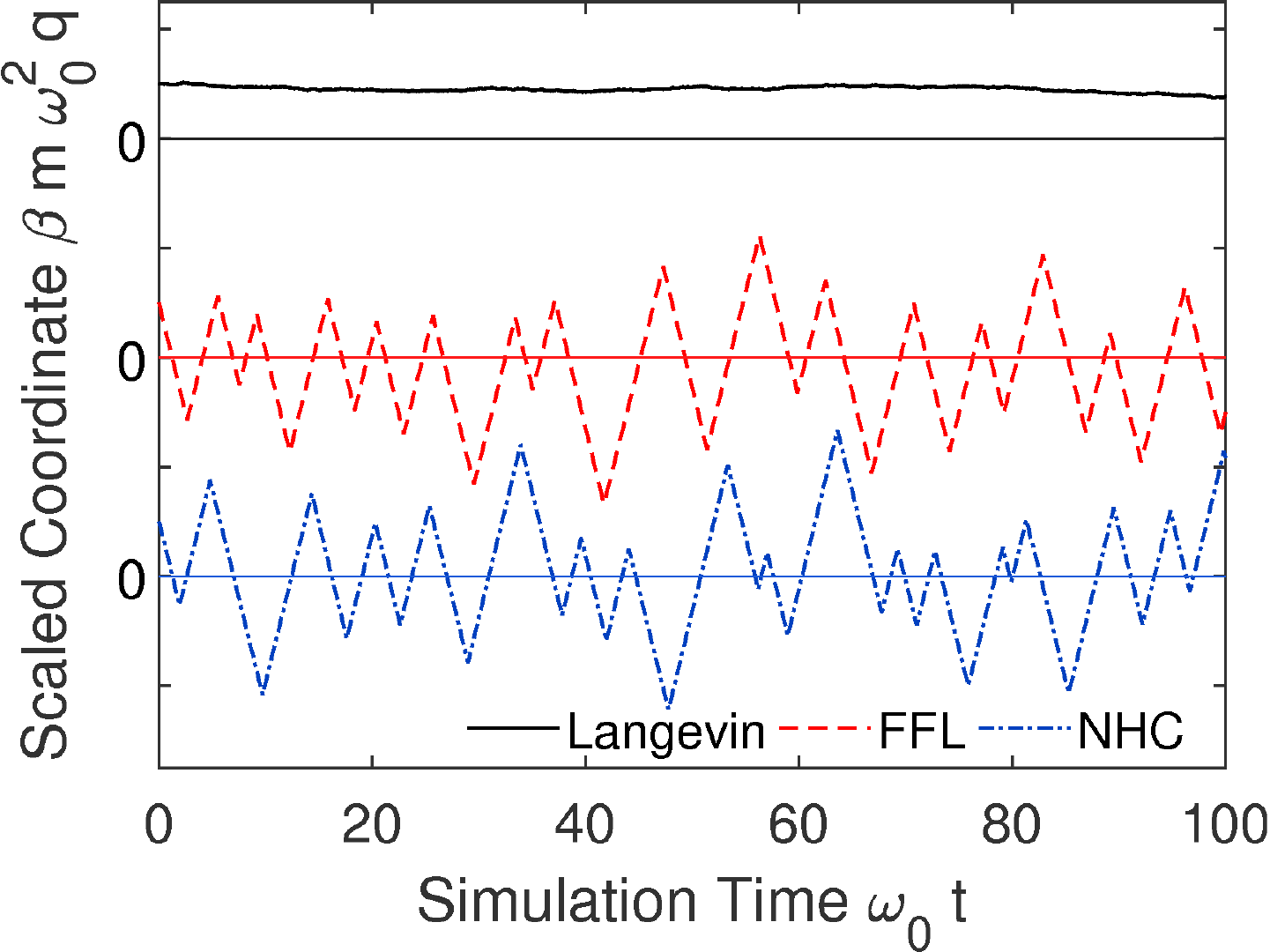}
\caption{\label{fig:trajectories}
Trajectories for a 1D harmonic oscillator with Langevin, FFL and NHC thermostats, with $\tau_{0}\omega_{0} = 10^{-4}$. The units are chosen such that $m = k_{\rm B} = 1$. In each case the calculation was started off with $q = -p = 1$.
}
\end{figure}

The dynamical effect of these three thermostats can be understood in more detail in Fig.~\ref{fig:trajectories}, which shows the trajectories of a harmonic oscillator with $\tau_{0}\omega = 10^{-4}$.
In this limit, plain Langevin thermostatting leads to strong overdamped behavior, and greatly hampers sampling efficiency. Both the FFL and the NHC allow the relevant configuration space to be sampled much more efficiently. 
The FFL and NHC thermostats generate similar - and rather peculiar - sawtooth trajectories. The rapid oscillation of the value of particle momentum leads to trajectories that, on the timescale of $1/\omega_0$ correspond to constant-average-velocity stretches separated by changes of direction induced by the potential. 
The similarity between the two schemes, as well as the equivalent performance, reinforces the notion that strongly-coupled NHC thermostats behave essentially like a more complicated version of their stochastic counterpart -- with the chain generating chaotic dynamics.

\subsection{Liquid Water}

Following the success of the FFL thermostat in improving the sampling efficiency in a simple 1D system, we next applied it to more complex, high-dimensional systems.
We ran simulations of equilibrated systems containing 216 molecules of q-TIP4P/F water\cite{Qwater} in a cubic simulation box with the experimental density of liquid water. The temperature was held constant at 298 K using the standard Langevin thermostat and the three generalizations of the FFL thermostat, for several values of $\tau_{0} = 1/\gamma$. For each value of the characteristic time, we ran 8 simulations of 10 ns and computed the potential energy correlation time $\tau_{V}$.
All simulations were run using the i-PI code.\cite{ipi}

\begin{figure}[h!]
\includegraphics[scale=0.5]{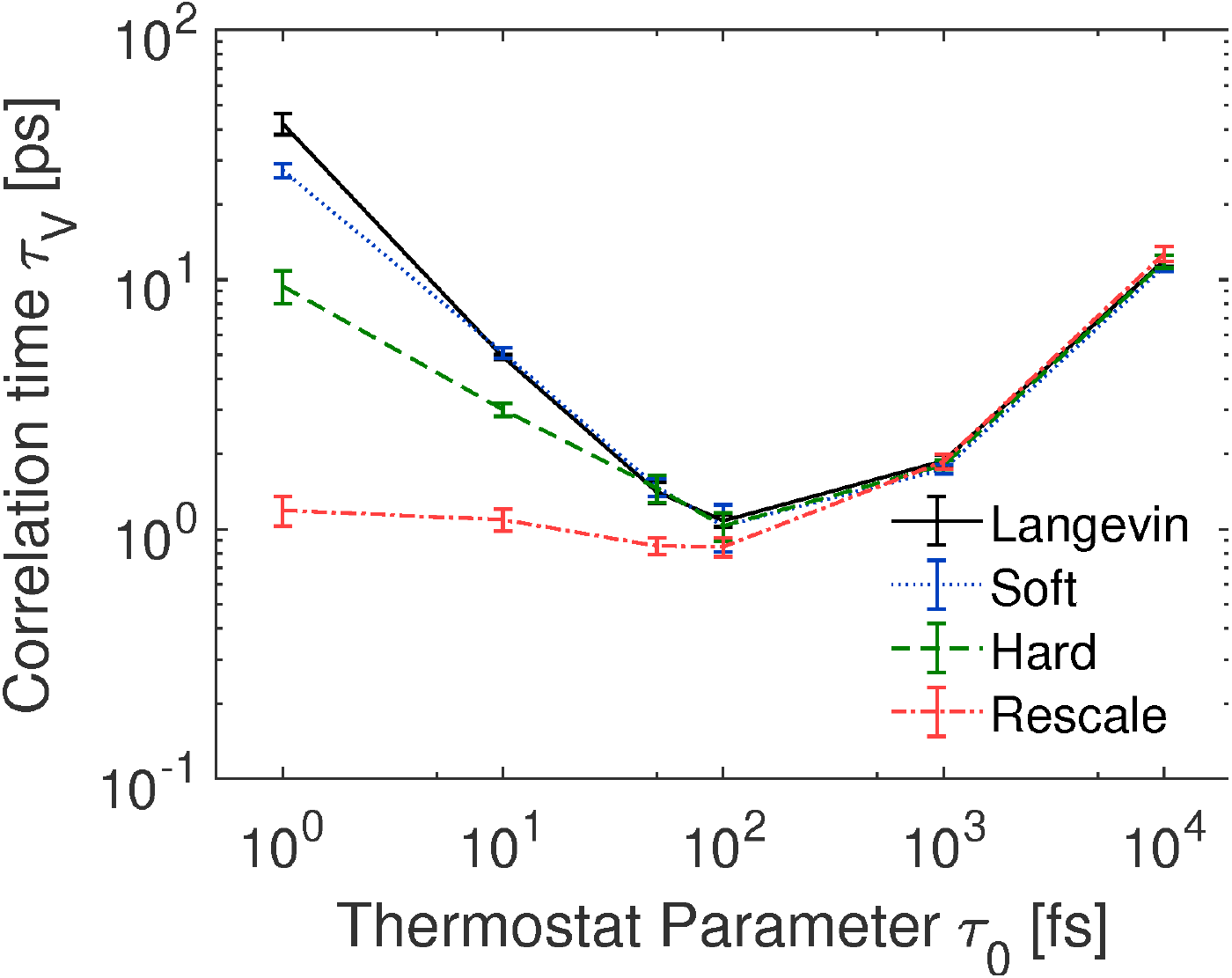}
\caption{Potential energy autocorrelation time $\tau_{V}$ for water using the standard Langevin thermostat (solid black curve), and the soft flip (dotted blue curve), hard flip (dashed green curve) and rescale flip (dash-dotted red curve) FFL thermostats. \label{fig:WatPE}}
\end{figure}

Fig.~\ref{fig:WatPE} shows $\tau_{V}$ as a function of $\tau_{0}$ for the four different types of thermostat.
Qualitatively, the standard Langevin equation shows the same behaviour as in Fig.~\ref{fig:Har} for the 1D harmonic oscillator, with a minimum correlation time at $\tau_{0} \simeq 100~\text{fs}$.
Both the hard and soft flipping algorithms show the same kind of trend, with a minimum at the same characteristic time. For low frictions they give the same $\tau_{V}$ as the standard algorithm, and for high frictions a slightly lower correlation time. However, for the soft flip this improvement is seen only at the highest friction and even for the hard flip the improvement is quite small.
The rescale flip, on the other hand, gives a correlation time that is qualitatively more similar to the 1D FFL in Fig.~\ref{fig:Har}: at high frictions, $\tau_{V}$ reaches a plateau, with a sampling efficiency almost two orders of magnitude greater than that of the standard Langevin thermostat.

A significant factor in determining the difference in results between the hard and soft flips and the rescale flip is that while the latter preserves the direction of the momentum of each particle, the former two methods do not; this means that after a few FFL steps the direction of the momentum could still reverse. Table~\ref{tab:diffusion} shows the self-diffusion coefficient $D$ of liquid water under the action of the different thermostats, in the underdamped and overdamped limits, and close to the critical damping.
Since,
\begin{equation}
D = \frac{1}{3}\int_{0}^{\infty} \left\langle \boldsymbol{v}(0)\cdot\boldsymbol{v}(t)\right\rangle\,{\rm d} t,
\end{equation}
is the integral of the autocorrelation function of the velocity, this gives an idea of the extent to which thermostatting changes the direction of momentum.
In the underdamped limit, all four thermostats give the same diffusion coefficient within the error bars, identical to its value for the q-TIP4P/F model in the NVE ensemble.\cite{Qwater}
At lower values of $\tau_{0}$, the standard Langevin and the soft and hard flip algorithms give very similar diffusion coefficients, while the rescale flip allows the system to diffuse much faster, avoiding the changes in momentum direction allowed by all of the other algorithms.
As shown in Ref.~\citenum{ceriotti2010}, however, for a highly-ergodic system such as liquid water the stochastic velocity rescaling thermostat~\cite{bussi_vresc2} offers the best performance, guaranteeing strong coupling without appreciable loss of diffusivity. 

\begin{table}[t]
\caption{\label{tab:diffusion} Self-diffusion coefficients of liquid water in the NVT ensemble at 298 K, for the standard Langevin thermostat and three types of FFL thermostat, at different values of the characteristic time $\tau_{0}$.}
\begin{tabular}{c | c c c}
\hline\hline
~~$\tau_{0}$ [fs]~~ & ~~10$^{0}$~~ & ~~10$^{2}$~~ & ~~10$^{4}$~~ \\
\hline
& \multicolumn{3}{c}{$D$ [\AA / ps$^{2}$]} \\
~~Langevin~~ & ~~0.038(0)~~ & ~~0.100(1)~~ & ~~0.192(3)~~ \\
~~Soft Flip~~ & ~~0.038(0)~~ & ~~0.098(1)~~ & ~~0.193(4)~~ \\
~~Hard Flip~~ & ~~0.031(0)~~ & ~~0.102(1)~~ & ~~0.190(2)~~ \\
~~Rescale Flip~~ & ~~0.083(0)~~ & ~~0.154(2)~~ & ~~0.192(3)~~ \\
\hline\hline
\end{tabular}
\end{table}

\subsection{Lennard-Jones Polymer}

Another system that shows a rich variety of behaviour is the Lennard-Jones polymer.\cite{LennardJP,LennardJHarP}
In this model, monomers interact with each other through a Lennard-Jones potential, $V(r) = 4 \epsilon \left[ (\sigma/r)^{12} - (\sigma/r)^{6}\right]$, and neighbouring monomers interact through a harmonic potential with force constant $k$ and equilibrium distance $r_{0}$. We take $\sigma = 2^{-1/6} r_{0}$ and $\epsilon = \frac{1}{72} k r_{0}^{2}$, so that the two potentials have the same equilibrium distance and the same second-derivative at this distance.

We used i-PI to perform simulations for a chain length of 128 monomers at a temperature of 1.9 $T^{\ast}$ (where an asterisk denotes Lennard-Jones units). This temperature is close to, but lower than, that of the globular-unwound transition.\footnote{As in Ref.~\onlinecite{LennardJHarP}, a transition temperature of around 2 $T^{\ast}$ was found by calculating the average potential energy of the polymer chain; Venkat Kapil, private communication.}
The proximity to the transition temperature means that convergence of the correlation times will be more challenging.
Once again, we performed 8 simulations with each value of $\tau_{0}$, with duration $8.6\times 10^{4}~t^{\ast}$.
The starting systems for these simulations were equilibrated as described in Ref.~\onlinecite{LennardJHarP}.

\begin{figure}[h!]
\includegraphics[scale=0.5]{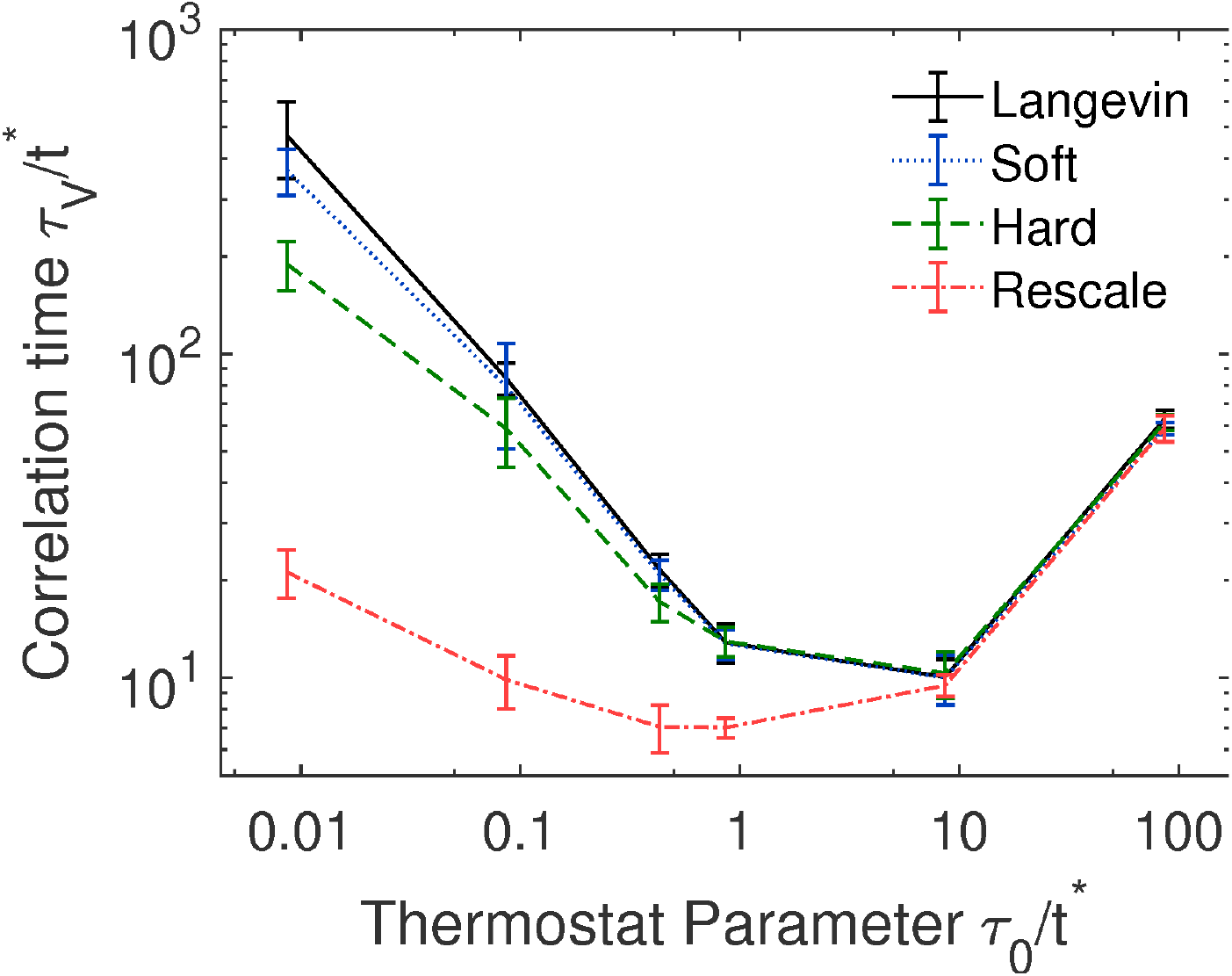}
\includegraphics[scale=0.5]{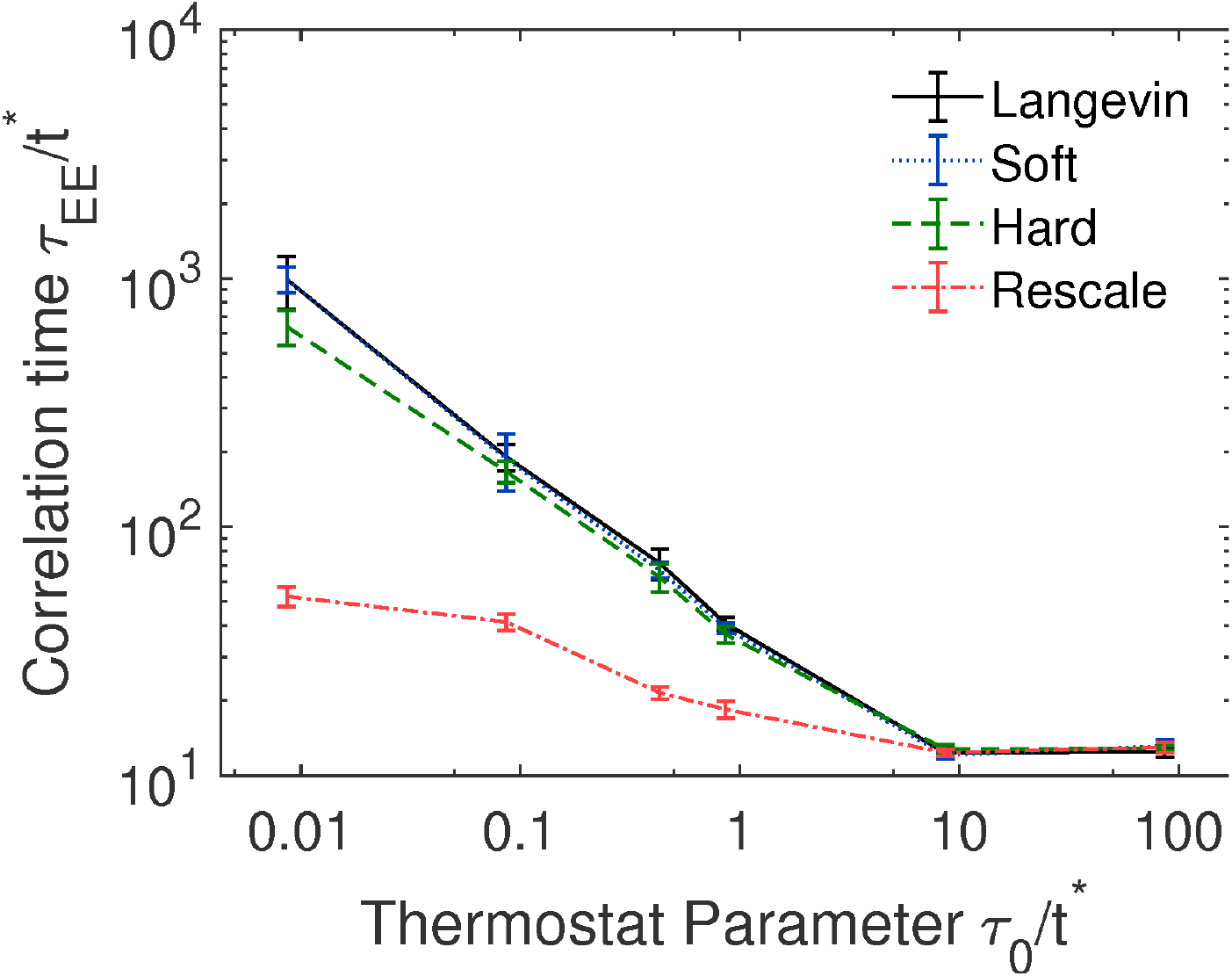}
\caption{\label{fig:LJ_times} Potential energy autocorrelation time $\tau_{V}$ (top panel) and end-to-end distance correlation time $\tau_{\rm EE}$ (bottom panel) for a Lennard-Jones polymer using the standard Langevin thermostat (solid black curve), and the soft flip (dotted blue curve), hard flip (dashed green curve) and rescale flip (dash-dotted red curve) FFL thermostats. All quantities are in LJ units.}
\end{figure}

The sampling efficiency was assessed using two correlation times: $\tau_{V}$ for the potential energy and $\tau_{\rm EE}$ for the end-to-end distance between the two ends of the chain. Fig.~\ref{fig:LJ_times} shows these two correlation times as a function of the characteristic time $\tau_{0}$.
For the potential energy correlation time, the standard Langevin equation shows the same qualitative behaviour as observed in previous results, as do the soft and hard flip algorithms. In this case, the soft flip provides no improvement over the Langevin thermostat, and the hard flip leads only to a very small improvement. As before, the rescale flip gives a significant decrease of correlation time -- in this case, two orders of magnitude -- but unlike previous results, it does not reach a plateau at the highest friction we have considered. 

The end-to-end distance correlation time $\tau_{\rm EE}$, on the other hand, gives quite different results: for all of the thermostats, the correlation time reaches a plateau at low frictions, giving more efficient thermostatting than at high frictions.
This reflects the fact that this distance is not a local variable, depending instead on the motion of the entire polymer, so that the local thermostatting we use here can significantly disturb the dynamics.
For the values of $\tau_{0}$ that give the most efficient behaviour, the FFL thermostats have no effect on the dynamics. However, for higher frictions the rescale flip does improve the sampling efficiency for the end-to-end distance, effectively increasing the range over which Langevin dynamics can be used to thermostat the system. The soft and hard flipping methods give no appreciable improvement over the standard Langevin equation.

\section{Discussion}\label{sec:discussion}

In all of the systems we have studied, the FFL thermostat has performed at least as well as the standard Langevin thermostat, and in the limit of high friction, gives sampling that is much more efficient.
This means that including momentum flips increases the range of $\gamma$ over which Langevin dynamics can be used to control the temperature of a system, and thus helps to ameliorate the major disadvantage of Langevin thermostatting compared to Nos{\'e}-Hoover.
In fact, this also helps to provide a more intuitive picture of the effect of a NHC thermostat: the coupling of multiple chains gives a chaotic dynamics that improves the ergodicity compared to the original NH thermostat, with the result that NHC dynamics are very similar to Langevin dynamics under the constraint that the sign of the momentum does not change due to the action of the thermostat (a result of the nonlinear form of the NHC equations of motion), but only due to the external force.

Although there is no unique generalization of the 1D FFL algorithm to multiple dimensions, the rescale flip, in which the direction of the momentum does not change, is probably the most natural one. There is a parallelism between the 1D FFL, the rescale-flipping algorithm and the SVR thermostat,\cite{VelocityRescaling}
in which constant-temperature sampling is obtained by an appropriate stochastic evolution of the system's kinetic energy.\cite{bussi_vresc2} This means that FFL dynamics is a local version of the SVR, without applying a correction
that allows the momentum to change sign.
This locality allows the FFL thermostat to more efficiently sample local properties, and thus to better equilibrate non-ergodic systems.
However, for more ergodic systems the global SVR thermostat is a better choice, as it will give effective thermostatting with very little disturbance of the system's dynamics.\cite{ceriotti2010}

While the rescale flip increases the sampling efficiency significantly compared to the standard Langevin equation, the soft and hard flipping methods lead to smaller effects. 
This is because these two algorithms do not preserve the direction of the momentum, so that after several thermostat steps the direction is able to reverse.
Unlike the rescale flip, these two methods have no real advantage over Langevin dynamics.

A key advantage of the FFL thermostat is its ease of implementation, either from scratch or as a modification of an existing Langevin dynamics algorithm.
In contrast, Nos{\'e}-Hoover chains involve a much more complex time-evolution, and thus more effort to implement.
Further, as noted in Ref.~\onlinecite{ceriotti2010}, unlike Langevin thermostats the NHC algorithm can add a significant overhead to calculations that use empirical forcefields.
These operational advantages, along with the improved sampling efficiency in the high-friction regime, mean that fast-forward Langevin dynamics provide a promising method for temperature control in molecular dynamics. 
An extension of this approach to the generalized Langevin equation thermostat is trivial, although it might not be obvious to recover the many analytical estimators that make it possible to fine-tune sampling~\cite{ceriotti2010} and dynamics~\cite{rossi2018}. 

\section{Conclusions}\label{sec:conclusions}

In this paper, we have introduced fast-forward Langevin dynamics, a modification of the traditional Langevin dynamics that improves the efficiency of sampling the canonical ensemble in the high-friction limit by flipping the momentum whenever the thermostat changes its direction, avoiding the over-damped behavior that would otherwise slow down exploration of configuration space.
This results in an algorithm akin to a local version of the stochastic velocity rescaling thermostat,
which is expected to improve sampling in non-ergodic systems.
By comparison with the NHC thermostat, we see that the additional steps we have added to Langevin dynamics remove their major disadvantage compared to Nos{\'e}-Hoover dynamics.
We also note a strong analogy in the overdamped limit between the chaotic motion of a system thermostatted by NHC and Brownian dynamics in which the direction of a particle's momentum does not change.

The simplicity of FFL dynamics, coupled with the improvement it gives over traditional Langevin dynamics, makes it a useful tool in atomistic modelling.
This algorithm could be easily extended to integrate generalized Langevin dynamics,\cite{alacarte} as well as to provide efficient sampling of path integral simulations.\cite{ceriotti2010}

\begin{acknowledgments}

The authors thank Venkat Kapil for useful discussions about the Lennard-Jones polymer simulations, and for critical reading of the manuscript.
This work was also supported by EPFL through the use of the facilities of its Scientific IT and Application Support Center. MC and DMW acknowledge support by the European Research Council under the European Union's Horizon 2020 research and innovation programme (grant agreement no. 677013-HBMAP). 
 
\end{acknowledgments}

\end{document}